\documentclass[num-refs, hidelinks]{nbdt-article}

\usepackage{siunitx}
\usepackage{float}

\papertype{Original Article}
\paperfield{Commentary on Generative Adversarial Collaboration: How do we generalize?}

\title{A roadmap to reverse engineering real-world generalization by combining naturalistic paradigms, deep sampling, and predictive computational models}

\author[1]{Peer Herholz}
\author[2,3]{Eddy Fortier}
\author[4,5]{Mariya Toneva}
\author[6]{Nicolas Farrugia}
\author[7,8]{Leila Wehbe}
\author[2,3]{Valentina Borghesani}

\affil[1]{ NeuroDataScience - ORIGAMI laboratory, The Neuro (Montreal Neurological Institute-Hospital), McGill University, Montréal, QC, Canada}
\affil[2]{Centre de Recherche de l'Institut Universitaire de Gériatrie de Montréal, Université de Montréal, Montréal, QC, Canada} 
\affil[3]{Département de Psychologie, Université de Montréal, Montréal, QC, Canada}
\affil[4]{Neuroscience Institute, Princeton, NJ, USA}
\affil[5]{Max Planck Institute for Software Systems, Saarbrücken, Germany}
\affil[6]{IMT Atlantique, Lab-STICC, UMR CNRS 6285, Brest, France}
\affil[7]{Machine Learning Department, Carnegie Mellon University, Pittsburgh, PA, USA}
\affil[8]{Neuroscience Institute, Carnegie Mellon University, Pittsburgh, PA, USA}

\corremail{herholz.peer@gmail.com or valentinaborghesani@gmail.com}

\runningauthor{Herholz et al.}

\begin{document}
\maketitle
\begin{abstract}
Real-world generalization, e.g., deciding to approach a never-seen-before animal, relies on contextual information as well as previous experiences. Such a seemingly easy behavioral choice requires the interplay of multiple neural mechanisms, from integrative encoding to category-based inference, \\ weighted differently according to the circumstances. 
Here, we argue that a comprehensive theory of the neuro-cognitive substrates of real-world generalization will greatly benefit from empirical research with three key elements. First, the ecological validity provided by multimodal, naturalistic \\ paradigms. Second, the model stability afforded by deep sampling. Finally, the statistical rigor granted by predictive modeling and computational controls. 
\keywords{generalization, deep sampling, naturalistic paradigm, predictive computational models}
\end{abstract}

\begin{figure}[H]
\centering
\includegraphics[scale=0.12]{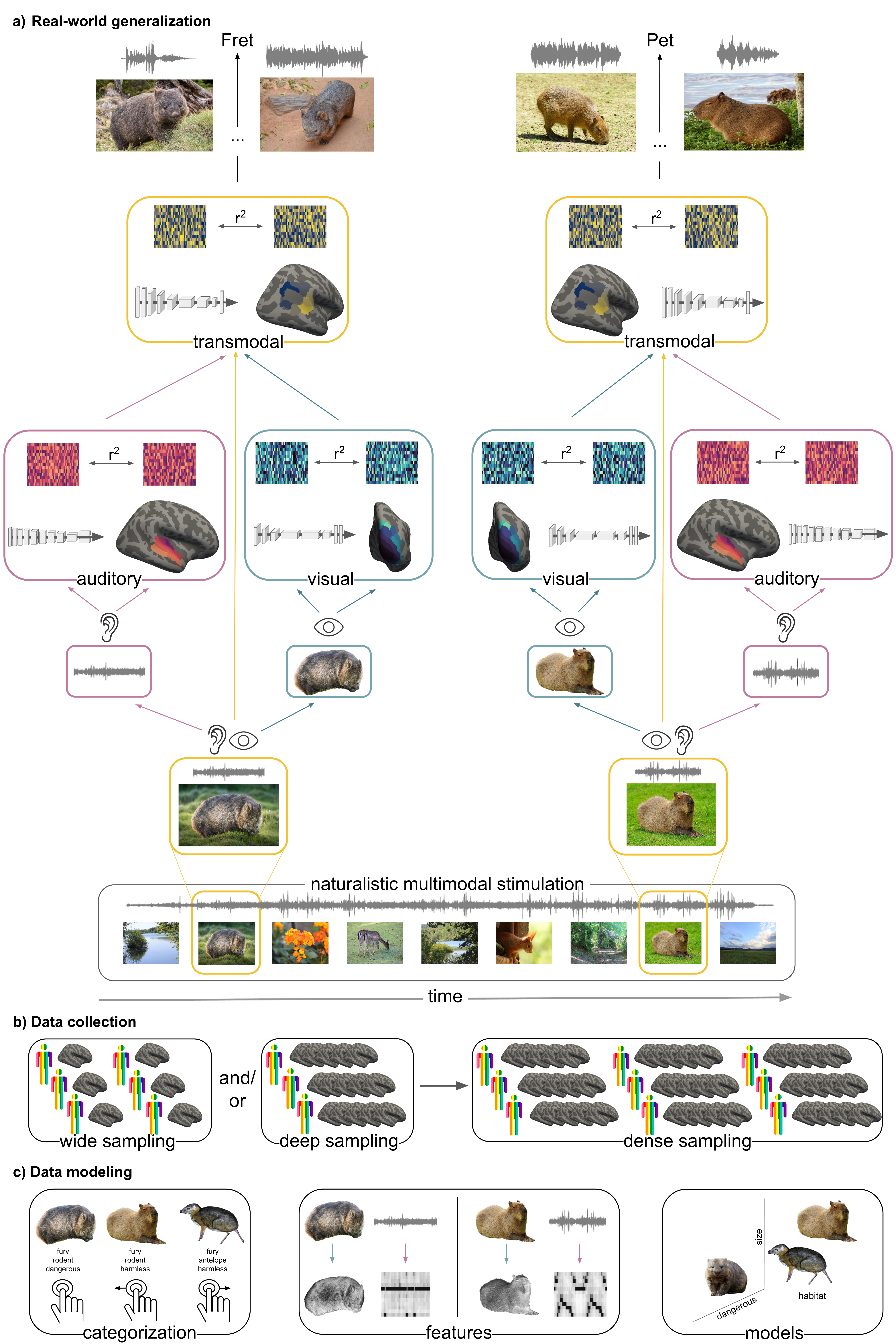}
\caption{Schematic representation of the discussed (a) real-world generalization process, (b) data collection procedures, and (c) modeling aspects. The capybara (South American rodent) and the wombat (Australian marsupial) share superficial visual features (e.g., quadrupedal, furry) yet only the latter can be dangerous for humans. Thus, they exemplify how a static image of an unknown animal would not be enough to choose the appropriate behavior (to pet or to fret?). The multimodal nature of our daily experiences forces us to include cross-modal paradigms enabling the comparison of generalization solutions across sensory systems, for example vision and audition (a). Moreover, with the ultimate goal of capturing both between- and within- subjects variability, we advocate for the need of collecting data from large samples as well as enough data from individual subjects (b). Finally, we stress the importance of combining behavioral paradigms, controlled features selection, and computational modeling (c).}
\label{overview}
\end{figure}

\newpage
Building and updating representations able to support complex, adaptive behaviors (e.g., \textit{Should I pet that unknown animal?}), is a critical feature of biological agents, and a yet unattainable goal for artificial ones. As reviewed by Taylor and colleagues \cite{taylor_how_2021}, notwithstanding the many different disciplines that have attempted to define generalization and describe its behavioral, cognitive, and neural underpinning, adjudicating between competing theoretical frameworks (e.g., memory integration vs. on-the-fly inference) is currently impossible. Building on the Authors' observation that progress will come from cross-laboratory, cross-disciplinary collaborations, we here suggest additional lines of empirical research that could pave the way to pivotal insights, namely: multimodal stimulation, naturalistic paradigms, deep sampling, and computational models.\\
\indent  Acknowledging the role of generalization in adaptive behavior necessarily means embracing the multimodal nature of our experiences. Approaching a never-seen-before short-legged, 40 inches long, furry animal might be a harmless or dangerous idea depending on whether you are facing a capybara or a wombat (Fig.~\ref{overview}a). While a static image would not be sufficient, the way they move, the sounds they make, the location of the encounter, would be more telling. Traditionally, empirical research focused on one sensory modality at a time, but nowadays a diverse spectrum of modalities can be the object of scientific inquiry. Crucially, advances in neuroimaging technology now enable multisensory stimulations, widening the scope of feasible studies (see for instance high quality fMRI auditory paradigms \cite{peelle_methodological_2014}). A better characterization of modality-specific networks and computations can thus be complemented with the exploration of how such features are integrated into the transmodal representations that ultimately lead to the behavioral choice (e.g., \textit{I am not frightened by that animal, I will pet it.}). Ultimately, cross-modal paradigms will allow us to test the limits of what generalizations can be inferred from individual modalities and what generalizations require the interaction of multiple modalities.\\
\indent  In real-world scenarios, generalization is enabled not only by contextual information (e.g., \textit{I am in Tasmania.}) but also by previous experiences: prior knowledge and beliefs are readily triggered and integrated, contributing to the prediction of possible outcomes \cite{clark_whatever_2013}. Moreover, generalization can operate on representations evoked in absence of any sensory stimulation, by internal, endogenous processes such as self-generated thoughts which only partially overlap with perception \cite{murphy_imagining_2019}, thus calling for dedicated experimental paradigms \cite{linke_flexible_2015}. Ultimately, to fully understand how real-world generalization works we will need to complement the highly controlled paradigms presented by Taylor and colleagues \cite{taylor_how_2021} with experimental conditions that are closer to real life’s problem-solving situations. According to basic methodology guidelines, the ability of a study to generalize to other contexts depend on its external validity \cite{ferguson_external_2004}. As defined by Andrade \cite[p. 999]{andrade_internal_2018}, the component of this validity that “examines whether the findings of a study can be generalized to naturalistic situations” is called ecological validity. In order to provide good ecological validity, stimuli must thus recreate as closely as possible a plausible scenario \cite{lewkowicz_concept_2001}.\\
\indent Multimodal, naturalistic datasets pose additional challenges to both data acquisition and analysis (Fig. ~\ref{overview}b). Moreover, the growing awareness of reproducibility issues has highlighted the importance of large sample sizes in neuroimaging studies \cite{turner_small_2018}. This is usually interpreted as a necessity that studies have a high number of participants (i.e., wide sampling). However, given the restrictions on total scan time and cost, a large number of participants also means a lower amount of data per participant. This leads to another reproducibility problem. As we know from statistical learning theory, it is essential to have a large number of data points to estimate models reliably \cite{gratton_editorial_2021}. With a fixed amount of resources, a strategy for building reproducible models and eventually learning robust between-subject differences across a population is to collect, as an initial step, a lot of data from a smaller number of participants (i.e., deep sampling). As this occurs in multiple labs across the world, we eventually will have a lot of data from a lot of participants (i.e., dense sampling). 

Once we have access to such rich datasets, one final important piece is computational modeling (Fig. ~\ref{overview}c). We argue that multivariate predictive modeling of brain recordings offers two key advantages. First, since it learns to associate brain activity to stimulus properties, it can help elucidate the representational content of brain areas that have been previously associated with different types of generalization. Second, it makes predictions on previously unseen stimuli \cite{mitchell_predicting_2008}, enabling the study of out-of-distribution generalization.. We further posit that predictive modeling is especially important when using the kind of ecologically valid stimuli that we advocate for. Ecologically valid stimuli offer enticing complexity but lack the precise controls of traditional neuroscience experiments, thereby making it more difficult to make scientific inferences. Recent work has proposed to isolate the main effect under investigation in naturalistic settings via computational controls \cite{toneva_combining_2020}. Computational controls model additional multivariate sources of variance in brain recordings and treat them as confounders in a post hoc analysis, akin to regressors in univariate analyses. For example, computational controls can be used to isolate the multivariate transmodal representation in the brain during movie watching from the multivariate visual and auditory sensory representations. Similarly to a controlled experiment, computational controls can only account for confounds that the researcher can operationalize and may not cover all intervening factors. Nevertheless, we believe that employing computational controls together with rich ecologically valid datasets will lead to more precise and generalizable scientific inferences.


To sum up, we argue that the quest for a comprehensive theory of the neuro-cognitive substrate of human real-world generalization ability will greatly benefit from  the adoption of (1) ecologically valid, multimodal, naturalistic paradigms, (2) deep sampling of both within and between subjects variability, and (3) multivariate predictive modeling and computational controls.

\section*{Acknowledgements}
The Authors would like to thank Dr. Pierre Bellec for his feedback on this commentary. V.B. is supported by a postdoctoral fellowship from the Institut de Valorisation des Données (IVADO) as well as research funding from the Courtois NeuroMod Project  (principal investigator Dr. Pierre Bellec, La Fondation Courtois). P.H. was supported in parts by funding from the Canada First Research Excellence Fund, awarded to McGill University for the Healthy Brains for Healthy Lives initiative, the National Institutes of Health (NIH) NIH-NIBIB P41 EB019936 (ReproNim), the National Institute Of Mental Health of the NIH under Award  Number  R01MH096906, a research scholar award from Brain Canada, in partnership with Health Canada, for the Canadian Open Neuroscience Platform initiative, as well as an Excellence Scholarship from Unifying Neuroscience and Artificial Intelligence - Québec. E.F. is supported by a master’s bursary from Courtois NeuroMod Project (principal investigator Dr. Pierre Bellec, La Fondation Courtois).

\section*{Conflict of interest}
None.

\printendnotes

\newpage
\renewcommand{\bibsection}{}
\section*{References}
\bibliography{CCN_GAC_Generalization_Commentary}

\end{document}